\documentclass{article}
\usepackage{amssymb,amsmath}
\usepackage{graphicx}
\usepackage{comment}

\newtheorem{defn}{\textbf{Definition}}
\newtheorem{thm}{\textbf{Theorem}}
\newtheorem{lem}{\textbf{Lemma}}
\newtheorem{prop}{\textbf{Proposition}}
\newtheorem{rem}{\textbf{Remark}}

\def\qed{\hfill \mbox{\rule[0pt]{1.5ex}{1.5ex}}}
\date{}
\title{A small-gain criterion for 2-contraction of large scale interconnected systems}
\author{David Angeli, Davide Martini, Giacomo Innocenti and Alberto Tesi
\thanks{David Angeli is with the Department of Electrical and Electronic Engineering, Imperial College London, London, SW7 2AZ, UK. {\textit{ d.angeli@imperial.ac.uk}} and with the Dept of Information Engineering of University of Florence; Giacomo Innocenti, Davide Martini and Alberto Tesi are with the Dept of Information Engineering of University of Florence.}
}
\begin{document}
\maketitle
\begin{abstract}
Despite modular conditions to guarantee stability for large-scale systems have been widely studied, few methods are available to tackle the case of networks with multiple equilibria. This paper introduces small-gain like sufficient conditions for $2$-contraction of large-scale interconnected systems on the basis of a family of upper-bounds to the $L_2$ gains that arise from the gains computed on individual channels of the second additive variational equation. Such a condition guarantee the $2$-additive compound of the system's Jacobian to be exponentially contractive, thus implying convergence towards equilibria of the system's solutions. The gains are obtained by solving  suitable Linear Matrix Inequalities. Three interconnected Thomas' systems are considered in order to illustrate the application of the theory and the degree of conservatism.
\end{abstract}
\section{Introduction}
The prediction of the long term behavior of nonlinear dynamical systems is a challenging and a hard topic that has caught the interest of the scientific community for a long time. This is also strongly related with assessment of stability and instability properties of solutions of a dynamical system. 
During the years, this topic has been tackled from different points of view, giving rise of many complementary approaches such as Lyapunov-based analysis \cite{sastry2013nonlinear}, Input-to-State Stability (ISS) \cite{sontag1995input}, passivity \cite{schaft1996l2}, monotonicity \cite{HirschSmith2005}, contraction theory \cite{slotinecontraction}, Incremental Stability \cite{angeliincremental}, or extreme stability \cite{yoshizawa}, just to name a few. In recent years, $k$-Contraction theory, based on the seminal paper by James Muldowney \cite{muldowney}, has gained interest (see \cite{kcontraction} and \cite{bar2023compound} for a recent survey on the topic). The idea relies on the connection between compound matrices and linear time-varying differential equations and, as a consequence, with nonlinear dynamical systems through the variational equation. 
In particular, the method imposes conditions on some matrix measures of the $k$-th additive compound of the Jacobian in order to quantify how 
$k$-dimensional perturbations with respect to initial conditions propagate along solutions. For $k=1$ we have the standard contraction property, while for $k=2$ we obtain Muldowney's conditions, which means that area perturbations are contracting along solutions in the state space of the system thus ruling out, on convex domains, the presence of periodic orbits. It has been also shown how it is possible to verify $k$-Contraction property without computing $k$-th order compound matrices \cite{dalin2023verifying}.  

Exploiting the algebra of compound matrices, it has allowed establishing a connection of the $2$-additive compound of the system's Jacobian with its Lyapunov exponents \cite{IEEEmartini},\cite{martini2023bounding}, providing a method to bound the maximal Lyapunov exponent and showing how it can be estimated by solving a certain number of Linear Matrix Inequalities. 
Furthermore, in \cite{angeli2023small}, the $2$-additive approach has been exploited to provide modular small-gain like conditions for $2$-contraction of a system constructed as the feedback interconnection of two subsystems. 
This latter approach allows reducing the size of the LMI problem to be solved, from $n \choose 2$, with $n$ dimension of the system, into three LMIs of dimension $n_1 \choose 2$, $n_2 \choose 2$ and $n_1 n_2$ respectively, where $n_1$ and $n_2$ are the dimensions of the two subsystems, $n_1+n_2=n$.

Consideration of a large number of interconnected subsystems, known in the literature as large-scale interconnected systems, has gained interest in the research community due to its widespread application in different fields, e.g. neural networks \cite{Hopfield}, dynamic nonlinear networks with coupled and multiterminal resistors, inductors, and capacitors \cite{chuagreen76}, biochemical networks \cite{Sontag}, large-scale dynamic systems \cite{largescaleSiljak}, swarm robotics \cite{brambilla2013swarm}. This is a challenging topic due to the intrinsic complexity of finding conditions for global stability or convergence of solutions, especially in cases where the overall system exhibits multistability. In particular, sufficient conditions often arise from a stability analysis. \\
Small-gain like criteria and input-output analysis have demonstrated to be a useful tool to tackle this problem. In \cite{vidyasagar1981input} the author investigate how to decompose and the well-posedness of large scale systems, before  treating the stability and instability problem with respect to different vector $L_p$ norms. While, in \cite{dashkovskiy2007iss}, authors provide small-gain like conditions for stability of ISS interconnected systems (relaxing the need for linear gain functions).

A further step in this direction is to consider a larger number of interconnected subsystems. Indeed, large-scale interconnected systems have gained a lot of interest in the research field due to their widespread application in different fields, e.g. neural networks \cite{Hopfield}, dynamic nonlinear networks containing coupled and multiterminal resistors, inductors, and capacitors \cite{chuagreen76}, biochemical networks \cite{Sontag}, large-scale dynamic systems \cite{largescaleSiljak}, swarm robotics \cite{brambilla2013swarm}. However, finding conditions for global stability or convergence of solutions is still a challenging problem, especially in cases where the overal system exhibits multistability. Among other approaches, small-gain like criteria and input-output analysis have demonstrated to be a useful tool to investigate stability of large-scale interconnected systems. For instance, a comprehensive input-output treatment of large scale systems can be found in \cite{vidyasagar1981input}, while \cite{dashkovskiy2007iss} provides small-gain like conditions for ISS-stability of interconnected systems.

In this paper, a system composed of more than two interconnected systems is considered. The goal is to extend the approach in \cite{angeli2023small} to this set-up, i.e., to provide small-gain like  conditions ensuring $2$-contraction of the overall system, thus enabling  convergence towards equilibria, i.e., in a multistable setting. AS in \cite{angeli2023small}, we look for conditions which can be verified by solving a finite number of LMIs, for which efficient software is available. The paper is structured as follows: Section \ref{sec:decomposition} shows how the $2$-additive compound matrix of a partitioned variational equation can be decomposed in a number of interconnected subsystems, whose equations can be written explicitly. For each subsystem a notion of $L_2$ gain is introduced in Section \ref{sec:L2gains}, while Section \ref{sec:SGlargescale} provides the main result on $2$-contraction of the overall system in terms of a modular small-gain condition. Section \ref{sec:proofMainRes} is devoted to the proof of the result, while an application example is discussed in Section \ref{sec:example}. Finally, some conclusion and final remarks end the paper in Section \ref{sec:concl}.
\section*{Notation}
Throughout the paper we adopt the following notations: 
\begin{itemize}
	\setlength{\itemsep}{0pt} \setlength{\parskip}{0pt}
	\item $\mathbb{N}$, $\mathbb{R}$, $\mathbb{C}$: sets of nonnegative integers, real numbers, complex numbers;
	\item $I_n$: $n\times n$ identity matrix;
	\item $A^T$: transpose of matrix $A$;
	\item $A \geq 0$ (resp. $A>0$): positive semidefinite (resp. definite) matrix $A$;
	\item ${\rm conv}(A_1, A_2, \ldots)$ : convex hull of matrices $A_1, A_2, \ldots$;
	\item $A^{(k)}$: $k$-multiplicative compound matrix of matrix $A$ ;
	\item $A^{[k]}$: $k$-additive compound matrix of matrix $A$;
	\item $[v_1, v_2, \ldots , v_m]$: matrix with column vectors \\$v_1, v_2, \ldots , v_m$;
	\item $A\otimes B$: Kronecker product of matrices $A$ and $B$;
	\item $A \oplus B$: Kronecker sum of matrices $A$ and $B$;
	\item $||\cdot||_2$: Euclidean $L_2$ norm;
	\item ${\rm card}(\mathcal{E})$: cardinality of the set $\mathcal{E}$;
 \item $\textrm{vec}(M)$: row vectorisation of rectangular matrix $M$;
 \item $\vec{X}$: strict upper row half-vectorisation of skew-symmetric matrix $X$.
\end{itemize}
\section{A decomposition of second additive compound linear equations}
\label{sec:decomposition}
Consider the following linear system of differential equations:
\begin{equation}
	\label{prototypevariational}
	\dot{\delta} (t) = A(t) \delta (t) 
\end{equation}
where $\delta \in \mathbb{R}^n$ is a column vector and $A \in \mathbb{R}^{n \times n}$ is a possibly time-varying matrix.
We consider the situation arising when $\delta$
can be partitioned in $N$ subvectors (of arbitrary heterogeneous dimensions), according to:
\begin{equation}
	\delta = [ \delta_1^T, \delta_2^T, \ldots, \delta_N^T]^T,  
\end{equation}
of dimensions $n_1, n_2, \ldots, n_N$ such that
$\sum_{k=1}^N n_k = n.$
Accordingly, matrix $A$ can be partitioned as:
\begin{equation}
	\label{matrixpartition}
	A = \left [ \begin{array}{cccc} A_{11} & A_{12} & \ldots & A_{1N} \\
		A_{21} & A_{22} &\ldots  & A_{2N}\\
		\vdots & \vdots & \ddots & \vdots \\
		A_{N1} & A_{N2} & \ldots &A_{NN} \end{array}
	\right ].
\end{equation}
Equations such as (\ref{prototypevariational}) arise
as variational equations of general nonlinear dynamical systems:
\begin{equation}
	\label{generalnonlin}
	\dot{x}(t) = f(x(t)),
\end{equation}
with state $x \in \mathbb{R}^n$ and defined through a $\mathcal{C}^1$ vector field $f(x): \mathbb{R}^n \rightarrow \mathbb{R}^n$. In such case $A(t)= \frac{ \partial f}{\partial x} (x(t))$, is the Jacobian evaluated along solutions of the nonlinear system (\ref{generalnonlin}).
An interesting notion of contraction for the study of variational equations of nonlinear systems, is the
so called $2$-contraction, as it allows to rule out existence of oscillatory behaviours (i.e. periodic, almost periodic or chaotic solutions) for the associated
system (\ref{generalnonlin}).
Such notion entails the exponential stability analysis of:
\begin{equation}
	\label{secondclassic}
	\dot{ \delta}^{(2)} (t) = A^{[2]} (t) \, \delta^{(2)} (t)
\end{equation}
where $A^{[2]} \in \mathbb{R}^{{n \choose 2} \times {n \choose 2}}$ denotes the so called second additive compound matrix, while $\delta^{(2)}$ is an auxiliary state vector in $\mathbb{R}^{n \choose 2}$.
Our goal is to establish $2$-contraction through a modular sufficient condition that exploits the partition of the state-space in analogy to what are normally referred to as large-scale small gain conditions in the context of standard Lyapunov asymptotic stability or contraction analysis. \\
To this end, we will make use of two types of vectorisation operators. The standard row vectorisation, which, for an arbitrary $m \times n$ rectangular matrix $X$ 
defines
\begin{equation}
	\label{skewvectorization}
	\textrm{vec}(X) := [ x_{11},x_{12}, \ldots, x_{1n}, x_{21}, x_{22}, \ldots, x_{2n}, \ldots, x_{m1}, \ldots, x_{mn} ]^T
\end{equation}
and a less common vectorisation operator which we denote
as $\vec{X}$, which only applies to skew-symmetric (square) matrices $X$ and is defined as follows:
\begin{equation}
	\label{skewvec}
	\vec{X}= [ x_{12}, x_{13}, \ldots, x_{1n}, x_{23},x_{24}, \ldots x_{2n}, \ldots, x_{(n-1)n}]^{T}.
\end{equation}
It is worth noting that the two operators in equations (\ref{skewvectorization}) and (\ref{skewvec}) are strictly related \cite{angeli_bifurcation}, \cite{marcus1960}. Indeed, for any skew-symmetric matrix $X\in\mathbb{R}^{n\times n}$ it holds
\begin{equation}
	\label{conversion}
	\textrm{vec}(X) = M_n \vec{X},
\end{equation}
where the matrix $M_n$ is given as:
\[ M_n = \sum_{1 \leq i \neq j \leq n} \textrm{sign}(j-i) e_{[(i-1)n + j]} e_{k(i,j)}^T  \]
with
\[ k(i,j) = |i-j|+ { n \choose 2 } - { {n+1 - \min \{i,j \}} \choose 2}. \]
Conversely, there exists a matrix $L_n \in \mathbb{R}^{{ n \choose 2 } \times n^2}$ such that the following holds:
\begin{equation}
	\label{conversion2}
	\vec{X} = L_n \textrm{vec} ( X ),
\end{equation}
where the matrix $L_n$ is given as:
\[ L_n = \sum_{1 \leq i < j \leq n}  e_{k(i,j)} e_{[(i-1)n + j]}^T.  \]
The adoption of operator (\ref{skewvec}) is particularly convenient due to the following known identification.
\begin{prop}
	Let $X(t)$ be an $n\times n$ skew-symmetric matrix. $X(t)$ is a solution of
	\begin{equation}
		\label{skewdifferential}
		\dot{X}(t) = A(t) X(t) + X(t) A(t)^T
	\end{equation}
	if and only if the corresponding vector $\vec{X}(t)$ fulfills:
	\begin{equation}
		\label{vectorisedskew}
		\dot{\vec{X}}(t) = A^{[2]} (t) \vec{X}(t).
	\end{equation}
\end{prop}
This result allows deriving a partition of $A^{[2]}$ (up to permutation of variables) on the basis of the original partition of $A$.
To this end, we partition a generic $n \times n$ skew symmetric matrix $X$ according to:
\begin{equation}
	\label{xpartition}
	X = \left [ \begin{array}{cccc} X_{11} & X_{12} & \ldots & X_{1N} \\
		X_{21} & X_{22} &\ldots  & X_{2N}\\
		\vdots & \vdots & \ddots & \vdots \\
		X_{N1} & X_{N2} & \ldots &X_{NN} \end{array}
	\right ],
\end{equation}
where each block $X_{ij}$ has the same dimensions as the corresponding block $A_{ij}$, and in addition we see that by virtue of skew-symmetry
$X_{ii}^T =  -X_{ii}$ for all $i =1, \ldots, N$, and 
$X_{ij}^T = - X_{ji}$ for all $i \neq j$.
\begin{lem}
	\label{a2structure}
	Consider the matrix differential equation (\ref{skewdifferential}) and assume that $X$ is partitioned as in (\ref{xpartition}),
	then for all $i=1,\ldots,N$ (such that $n_i \geq 2$) the following equation holds for the vectorised diagonal block of $X$:
	\begin{equation}
		\label{diagonalblocks}
		\dot{ \vec{X}}_{ii}(t) 
		= A_{ii}^{[2]}  \vec{X}_{ii} (t)
		+ \sum_{k < i} B_{ik} \textrm{vec} (X_{ki} (t) )
		+ \sum_{k>i} B_{ik} \textrm{vec} (X_{ik} (t) )
	\end{equation}
	where
	\begin{equation}
		\label{eq:Bik}
		B_{ik} = \left \{ 
		\begin{array}{rl}
			L_{n_i} [ (A_{ik} \otimes I_{n_i} ) -
			(I_{n_i} \otimes A_{ik} ) Q_{n_i,n_k} ] &
			\textrm{if } k<i \\
			L_{n_i} [ ( I_{n_i} \otimes A_{ik} ) - (A_{ik} \otimes I_{n_i} ) Q_{n_i,n_k} ] & \textrm{if }k > i
		\end{array} \right . 
	\end{equation}
	and the matrix $Q_{n_i,n_j}$ is defined as
	\[ Q_{n_i,n_j} = \sum_{h=1}^{n_i} \sum_{k=1}^{n_j} e_{[(k-1)n_i + h]} e_{[(h-1) n_j + k]}^T  \] 
	and converts row vectorisation to column vectorisation, viz. \\$\textrm{vec} ( X_{ij}^T ) = Q_{n_i,n_j} \textrm{vec} (X_{ij} )$.
	Moreover, for all $1 \leq i<j \leq N$ the following equations hold for the vectorised blocks of $X$ above the diagonal:
	\begin{eqnarray}
		\label{offdiagonalblocks}
		\notag\textrm{vec}(\dot{X}_{ij} ) &=& (A_{ii} \oplus A_{jj} ) 
		\textrm{vec} ( X_{ij} ) + G^1_{ij} \vec{X}_{ii}
		+ G^2_{ij} \vec{X}_{jj} + \sum_{i \neq k < j}
		H^{ij}_{kj} \textrm{vec} (X_{kj} )
		\\&+& \sum_{k > j} H^{ij}_{kj} \textrm{vec} (X_{jk} )
		+ \sum_{k<i} H^{ij}_{ki} \textrm{vec} (X_{ki} )
		+ \sum_{j \neq k > i} H^{ij}_{ki} \textrm{vec}( X_{ik} ) \nonumber
	\end{eqnarray}
	where:
	\begin{align}
		\label{eq:Gik}
		\notag &G^1_{ij} = (I_{n_i} \otimes A_{ji} ) \, M_{n_i} \\
		&G^2_{ij} = ( A_{ij} \otimes I_{n_j} ) M_{n_j}, 
	\end{align}
	and
	\begin{align}
		\label{eq:H}
		\notag &H^{ij}_{kj} = \left \{ \begin{array}{rl} ( A_{ik} \otimes I_{n_j} ) & \textrm{if } k<j \\
			- (A_{ik} \otimes I_{n_j} ) Q_{n_j,n_k} & \textrm{if } k>j 
		\end{array}
		\right .\\
		&H^{ij}_{ki} = \left \{
		\begin{array}{rl}
			(I_{n_i} \otimes A_{jk} ) & \textrm{if } k>i \\
			- (I_{n_i} \otimes A_{jk} ) Q_{n_k,n_i} & \textrm{if } k<i
		\end{array}
		\right .
	\end{align}
\end{lem}
\emph{Proof.}
Computing the block-partitioned expression of $\dot{X}$ according to
\begin{align*} 
	&\dot{X} =\\& \left [ \begin{array}{ccc} A_{11} & \dots &  A_{1N}\\ \vdots & \ddots & \vdots \\ A_{N1} & \cdots & A_{NN}\end{array} \right ] \,
	\left [ \begin{array}{ccc} X_{11} & \dots &  X_{1N}\\ \vdots & \ddots & \vdots \\ X_{N1} & \cdots & X_{NN}\end{array} \right ]
	+ \left [ \begin{array}{ccc} X_{11} & \dots &  X_{1N}\\ \vdots & \ddots & \vdots \\ X_{N1} & \cdots & X_{NN}\end{array} \right ] \, \left [ \begin{array}{ccc} A_{11} & \dots &  A_{1N}\\ \vdots & \ddots & \vdots \\ A_{N1} & \cdots & A_{NN}\end{array} \right ]^T=\\&\left [ \begin{array}{ccc} A_{11} & \dots &  A_{1N}\\ \vdots & \ddots & \vdots \\ A_{N1} & \cdots & A_{NN}\end{array} \right ] \,
	\left [ \begin{array}{ccc} X_{11} & \dots &  X_{1N}\\ \vdots & \ddots & \vdots \\ X_{N1} & \cdots & X_{NN}\end{array} \right ]
	+ \left [ \begin{array}{ccc} X_{11} & \dots &  X_{1N}\\ \vdots & \ddots & \vdots \\ X_{N1} & \cdots & X_{NN}\end{array} \right ] \, \left [ \begin{array}{ccc} A_{11}^T & \dots &  A_{N1}^T\\ \vdots & \ddots & \vdots \\ A_{1N}^T & \cdots & A_{NN}^T\end{array} \right ],
\end{align*}
the dynamics of $\dot{X}$ can be written in the following synthetic form:
\begin{align}
	\label{eq:Xii}
	\notag	&\dot{X}_{ii} = A_{ii}\,X_{ii} + X_{ii}\,A_{ii}^T + \sum\limits_{k<i}\left(A_{ik}\,X_{ki} +X_{ik}\,A_{ik}^T\right) + \sum\limits_{k>i}\left(X_{ik}\,A_{ik}^T +A_{ik}\,X_{ki}\right) \vspace{0.2cm}\\&=A_{ii}\,X_{ii} + X_{ii}\,A_{ii}^T + \sum\limits_{k<i}\left(A_{ik}\,X_{ki} -X_{ki}^T\,A_{ik}^T\right) + \sum\limits_{k>i}\left(X_{ik}\,A_{ik}^T -A_{ik}\,X_{ik}^T\right),\\
	\label{eq:Xij}
		\notag&\dot{X}_{ij} = A_{ii}\,X_{ij} + X_{ij}\,A_{jj}^T + A_{ij}\,X_{jj} + X_{ii}\,A_{ji}^T + \sum\limits_{i\neq k<j} A_{ik}\,X_{kj} + \sum\limits_{k>j} A_{ik}\,X_{kj} \\\notag&+ \sum\limits_{k<i} X_{ik}\,A_{jk}^T + \sum\limits_{j\neq k>i} X_{ik}\,A_{jk}^T =
		A_{ii}\,X_{ij} + X_{ij}\,A_{jj}^T + A_{ij}\,X_{jj} + X_{ii}\,A_{ji}^T \vspace{0.2cm}\\&+ \sum\limits_{i\neq k<j} A_{ik}\,X_{kj} - \sum\limits_{k>j} A_{ik}\,X_{jk}^T - \sum\limits_{k<i} X_{ki}^T\,A_{jk}^T + \sum\limits_{j\neq k>i} X_{ik}\,A_{jk}^T.
\end{align}
Applying the operator in (\ref{skewvec}) to both sides of equation (\ref{eq:Xii}) and exploiting the row vectorisation identity 
$\textrm{vec}(AXB^T) = (A \otimes B) \textrm{vec}(X)$, we obtain:
\begin{align*}
	&\dot{\vec{X}}_{ii} = \overrightarrow{\left(A_{ii}\,X_{ii} + X_{ii}\,A_{ii}^T\right)} + \sum\limits_{k<i}\overrightarrow{\left(A_{ik}\,X_{ki} -X_{ki}^T\,A_{ik}^T\right)} + \sum\limits_{k>i}\overrightarrow{\left(X_{ik}\,A_{ik}^T -A_{ik}\,X_{ik}^T\right)} \\&=A_{ii}^{[2]}\,\vec{X}_{ii} 
	+ \sum\limits_{k<i}L_{n_i}{\rm vec}\left(A_{ik}\,X_{ki} -X_{ki}^T\,A_{ik}^T\right)  + \sum\limits_{k>i}L_{n_i}{\rm vec}\left(X_{ik}\,A_{ik}^T -A_{ik}\,X_{ik}^T\right)\\&=A_{ii}^{[2]}\,\vec{X}_{ii} + \sum\limits_{k<i}L_{n_i}\,{\rm vec}(A_{ik}\,X_{ki}) -\sum\limits_{k<i}L_{n_i}\,{\rm vec}(X_{ki}^T\,A_{ik}^T) + \sum\limits_{k>i}L_{n_i}\,{\rm vec}(X_{ik}\,A_{ik}^T) \\&- \sum\limits_{k>i}L_{n_i}\,{\rm vec}(A_{ik}\,X_{ik}^T)=	A_{ii}^{[2]}\,\vec{X}_{ii} + \sum\limits_{k<i}L_{n_i}\,(A_{ik}\otimes I_{n_i})\,{\rm vec}(X_{ki}) \\&- \sum\limits_{k<i}L_{n_i}\,(I_{n_i} \otimes A_{ik})\,{\rm vec}(X_{ki}^T) + \sum\limits_{k>i}L_{n_i}\,(I_{n_i}\otimes A_{ik})\,{\rm vec}(X_{ik}) \\&- \sum\limits_{k<i}L_{n_i}\,(A_{ik}\otimes I_{n_i})\,{\rm vec}(X_{ik}^T).\\
\end{align*}
Then, using of matrix $Q_{n_i,n_j}$, which converts row vectorisation to column vectorisation, viz. ${\rm vec}(X_{ij}^T) = Q_{n_i,n_j} {\rm vec}(X_{ij})$, we obtain:
\begin{align*}
	\dot{\vec{X}}_{ii} &= A_{ii}^{[2]}\,\vec{X}_{ii} + \sum\limits_{k<i}L_{n_i}\,[(A_{ik}\otimes I_{n_i}) - (I_{n_i} \otimes A_{ik})\,Q_{n_i,n_k}]\,{\rm vec} (X_{ki})\vspace{0.2cm}\\& + 
	\sum\limits_{k>i}L_{n_i}\,[(I_{n_i}\otimes A_{ik})- (A_{ik}\otimes I_{n_i})\,Q_{n_i,n_k}]\,{\rm vec}(X_{ik}),
\end{align*}
thus proving (\ref{diagonalblocks})-(\ref{eq:Bik}). Next, taking the operator in (\ref{skewvectorization}) in both sides of equation (\ref{eq:Xij}) and applying the linearity property of the operator, yields:
\begin{align*}
	&{\rm vec}(\dot{X}_{ij}) = {\rm vec}(A_{ii}\,X_{ij}) + {\rm vec}(X_{ij}\,A_{jj}^T) + {\rm vec}(A_{ij}\,X_{jj}) + {\rm vec}(X_{ii}\,A_{ji}^T) \\&+ \sum\limits_{i\neq k<j} {\rm vec}(A_{ik}\,X_{kj}) - \sum\limits_{k>j} {\rm vec}(A_{ik}\,X_{jk}^T) - \sum\limits_{k<i} {\rm vec}(X_{ki}^T\,A_{jk}^T) \\&+ \sum\limits_{j\neq k>i} {\rm vec}(X_{ik}\,A_{jk}^T) =
	(A_{ii}\otimes I_{n_j})\,{\rm vec}(X_{ij}) + (I_{n_i}\otimes A_{jj})\,{\rm vec}(X_{ij}) \\&+ (A_{ij}\otimes I_{n_j})\,{\rm vec}(X_{jj}) + (I_{n_i}\otimes A_{ji})\,{\rm vec}(X_{ii}) + \sum\limits_{i\neq k<j} (A_{ik}\otimes I_{n_j})\,{\rm vec}(X_{kj}) \\&- \sum\limits_{k>j} (A_{ik}\otimes I_{n_j})\,{\rm vec}(X_{jk}^T) - \sum\limits_{k<i} (I_{n_i}\otimes A_{jk})\,{\rm vec}(X_{ki}^T) \\&+ \sum\limits_{j\neq k>i} (I_{n_j}\otimes A_{jk}^T)\,{\rm vec}(X_{ik}) =
	(A_{ii}\oplus A_{jj})\,{\rm vec}(X_{ij}) + (A_{ij}\otimes I_{n_j})\,M_{n_j}\,\vec{X}_{jj} \\&+ (I_{n_i}\otimes A_{ji})\,M_{n_i}\,\vec{X}_{ii}+ \sum\limits_{i\neq k<j} (A_{ik}\otimes I_{n_j})\,{\rm vec}(X_{kj}) - \sum\limits_{k>j} (A_{ik}\otimes I_{n_j})\,{\rm vec}(X_{jk}^T) \\&- \sum\limits_{k<i} (I_{n_i}\otimes A_{jk})\,{\rm vec}(X_{ki}^T) + \sum\limits_{j\neq k>i} (I_{n_j}\otimes A_{jk}^T)\,{\rm vec}(X_{ik})
\end{align*}
Finally, exploiting the matrix $Q_{n_i,n_j}$ we get:
\begin{align*}
	{\rm vec}(\dot{X}_{ij}) &= (A_{ii}\oplus A_{jj})\,{\rm vec}(X_{ij}) + (A_{ij}\otimes I_{n_j})\,M_{n_j}\,\vec{X}_{jj} + (I_{n_i}\otimes A_{ji})\,M_{n_i}\,\vec{X}_{ii}\\&+ \sum\limits_{i\neq k<j} (A_{ik}\otimes I_{n_j})\,{\rm vec}(X_{kj}) 
	- \sum\limits_{k>j} (A_{ik}\otimes I_{n_j})\,Q_{n_j,n_k}\,{\rm vec}(X_{jk})\\& - \sum\limits_{k<i} (I_{n_i}\otimes A_{jk})Q_{n_k,n_i}\,{\rm vec}(X_{ki}) + \sum\limits_{j\neq k>i} (I_{n_j}\otimes A_{jk}^T){\rm vec}(X_{ik}).\\
	&\qquad\qquad\qquad\qquad\qquad\qquad\qquad
	\qquad\qquad\quad\qquad\qquad\qquad\quad\quad\qed
\end{align*}
Lemma \ref{a2structure} introduces a decomposition of $\vec{X}$, (or equivalently of the auxiliary $\delta^{(2)}$ variable in equation (\ref{secondclassic})), in terms of the vectors $\textrm{vec} (X_{ij})$, for $1 \leq i < j \leq N$ and $\vec{X}_{kk}$, for $1 \leq k \leq N$. 
Moreover, it highlights the interactions among such variables as dictated by the dynamics of the second additive variational equation. \\

\section{$L_2$ gains for linear time-varying systems}
\label{sec:L2gains}

To derive a modular sufficient condition for stability of such interconnected system we rely on the notion of $L_2$ gain.
\begin{defn}
	For a linear system:
	\begin{equation}
		\label{diffincl}
		\dot{\delta} (t) = A(t) \delta(t) + B(t) u(t)     
	\end{equation}
	with $A(t)$ and $B(t)$ continuous time-dependent matrices of compatible dimension, belonging to some bounded
	set $\Omega \subset \mathbb{R}^{n \times n} \times \mathbb{R}^{n \times m}$, we say that $\gamma$ is an upper-bound to the $L_2$ gain if there exists $M>0$ such that following inequality is fulfilled:
	\begin{equation}
		\label{l2defi}
		\int_0^{+\infty} |\delta(t)|^2 dt \leq \gamma^2 \int_0^{+\infty} |u(t)|^2 dt + M |\delta(0)|^2 
	\end{equation}
	for all $\delta(0) \in \mathbb{R}^n$, for all measurable input signals $u(\cdot)$ and all solutions of (\ref{diffincl})
	with $[A(t),B(t)] \in \Omega$.
\end{defn}
\begin{rem}
	\label{l2LMI}
	It is well-known that a sufficient condition to validate $\gamma$ as an upper-bound to the $L_2$ gain is, in the case of $\Omega = \textrm{co} \{ [A_i,B_i]: 1 \leq i \leq Q \}$ through satisfaction of the following LMI conditions for some symmetric positive definite $P \in \mathbb{R}^{n \times n}$:
	\begin{equation}
		\left [ \begin{array}{cc}   A_i^T P + P A_i + I & P B_i \\ B_i^T P & - \gamma^2 I_m \end{array} \right ] \leq 0, \qquad 1 \leq i \leq Q.
	\end{equation}
\end{rem}
Due to the complicated nature of equations (\ref{diagonalblocks}) and (\ref{offdiagonalblocks}) it is convenient to introduce $L_2$ gains for systems whose input is partitioned into multiple separate channels ($R$ in this case).
\begin{defn}
	For a linear system:
	\begin{equation}
		\label{diffinclMI}
		\dot{\delta} (t) = A(t) \delta(t) + \sum_{k=1}^{R} B_k(t) u_k(t)     
	\end{equation}
	with $A(t)$ and $B_1(t), \ldots, B_R(t)$ continuous time-dependent matrices of compatible dimension, belonging to some bounded
	set $\Omega \subset \mathbb{R}^{n \times n} \times \prod_{k=1}^R \mathbb{R}^{n \times m_k}$, we say that $\gamma_k$, $k=1,\ldots R$ are a family of upper-bounds to the $L_2$ gains if there exists $M>0$ such that following inequality is fulfilled:
	\begin{equation}
		\label{l2defiMI}
		\int_0^{+\infty} |\delta(t)|^2 dt \leq \left ( \sum_{k=1}^R \gamma_k^2 \int_0^{+\infty} |u_k(t)|^2 dt  \right )+ M |\delta(0)|^2 
	\end{equation}
	for all $\delta(0) \in \mathbb{R}^n$, for all measurable input signals $u(\cdot)$ and all solutions of (\ref{diffincl})
	with $[A(t),B_1(t), \ldots,B_R(t)] \in \Omega$.
\end{defn}
\begin{rem}
	\label{l2LMI2}
	Similarly to the case of a single input channel, a sufficient condition to validate $\gamma_k$s  as a family of upper-bounds to the $L_2$ gains is, in the case of $\Omega = \textrm{co} \{ [A_i,B_1^i, B_2^i, \ldots, B_R^i]: 1 \leq i \leq Q \}$ through satisfaction of the following LMI conditions for some symmetric positive definite $P \in \mathbb{R}^{n \times n}$:
	\begin{equation}
		\label{bigLMI}
		\left [ \begin{array}{ccccc}   A_i^T P + P A_i + I & P B_1^i & P B_2^i & \ldots & P B_R^i \\ {B_1^i}^T P & - \gamma_1^2 I_{m_1} & 0 & \ldots & 0 \\ {B_2^i}^T P & 0 & - \gamma_2^2 I_{m_2} & \ddots & \vdots \\
			\vdots & \vdots & \ddots & \ddots & 0 \\ {B_R^i}^T P & 0 & \ldots& 0 & - \gamma_R^2 I_{m_R} \end{array} \right ] \leq 0, \qquad 1 \leq i \leq Q.
	\end{equation}
\end{rem}
While the condition highlighted in Remark \ref{l2LMI2} could in principle be adopted to compute a family of $L_2$ gains, it has two drawbacks. First, the size of the LMI grows with the size of the system considered and, in particular, this growth could be quadratic in $n$ or $N$. Additionally, while for a single gain one would normally minimize the value of the upper-bound in order to find the tightest possible characterization of systems' solutions, there is no natural way to do this in the case of multiple gains, i.e. a Pareto front of upper-bounds is normally encountered, leaving us the difficult task of deciding how to weight the different gains and which input channels to prioritize. \\

As an alternative, we propose a simple Lemma which allows to compute a family of upper-bounds to the $L_2$ gains on the basis of the gains computed on individual channels, and exploiting, to this end, the superposition principle of linear (time-varying) systems.
\begin{lem}
	\label{gainrescale}
	Consider the linear system in equation (\ref{diffinclMI}).
	Assume that for each $k=1, \ldots, R$ an upper-bound to the $L_2$-gain from $u_k$ to $\delta$ is known, viz.
	there is $\gamma_k \geq 0$ such that, for the system:
	\begin{equation}
		\label{deltaequationi}
		\dot{\delta}_k (t) = A(t) \delta_k (t) +  B_k(t) u_k(t),  
	\end{equation}
	it holds:
	\[  \int_0^{+ \infty} |\delta_k(t)|^2 \, dt \leq \gamma_k^2 \int_0^{+ \infty} |u_k(t)|^2\, dt + M_k |\delta_k(0)|^2  \]   
	for some sufficiently large $M_k>0$ and all $\delta_k(0) \in \mathbb{R}^n$, and regardless of the time-varying matrices $A(t)$ and $B_k(t)$.
	Then, for $M$ sufficiently large it holds:
	\[   \int_0^{+ \infty} |\delta(t)|^2 \, dt     
	\leq \sum_{k=1}^R R \gamma_k^2 \int_0^{+ \infty} |u_k(t)|^2  \, dt + 
	M |\delta(0)|^2, \]
	viz. $\sqrt{R} \gamma_k$ is a family of upper bounds to the $L_2$ gain of system (\ref{diffinclMI}). 
\end{lem}
\emph{Proof.}
To see this, it is enough to remark that defining
$\delta(t) = \sum_{k=1}^R \delta_k(t)$ we see it is a solution of (\ref{diffinclMI}), $\delta_k(t)$, $k=1, \ldots, R$, is a solution of (\ref{deltaequationi}).
Moreover, it can be verified that
\[    |\delta(t)|^2=\delta^T(t) \delta(t) = \left ( \sum_{k=1}^R \delta_k(t) \right )^T \left ( \sum_{k=1}^R \delta_k(t) \right ) \]
\[ \quad = \sum_{k=1}^R \delta_k(t)^T \delta_k(t) + \sum_{ k \neq l } \delta_k(t)^T \delta_l(t) 
\]
\[ \qquad \leq \sum_{k=1}^R |\delta_k(t)|^2  + \sum_{k \neq l} [ |\delta_k(t)|^2  + |\delta_l(t)|^2 ] / 2 = R \sum\limits_{k=1}^R |\delta_k(t)|^2 \]
Taking integrals of the above inequality yields:
\[
\int_0^{+ \infty} |\delta(t)|^2 \, dt \leq  R \sum_{k=1}^R \int_0^{+\infty} |\delta_k(t)|^2 \, dt 
\]
\[ \qquad \leq   \sum_{k=1}^R R \gamma_k^2 \int_0^{+ \infty} |u_k(t)|^2 \, dt
+ R \left ( \sum_{k=1}^R M_k \right )  |\delta (0)|^2. \]
This proves the Lemma with $M=  R \left ( \sum_{k=1}^R M_k \right )$.
\begin{rem}
	While Lemma \ref{a2structure} proves that $\sqrt{R} \gamma_k$, for $k=1, \ldots, R$ is a family of upper-bounds to the $L_2$ gain of system (\ref{diffinclMI}) it is not true in general that such gains can be validated through an LMI of the type (\ref{bigLMI}), even if the individual $\gamma_k$s are known to fulfill single channel inequalities of the type:
	\begin{equation}
		\label{singlechannelLMI}
		\left [ \begin{array}{cc}   A_i^T P_k + P_k A_i + I & P_k B^i_k \\ {B_k^i}^T P_k & - \gamma_k^2 I_m \end{array} \right ] \leq 0, \qquad 1 \leq i \leq Q.
	\end{equation}
	In other words, there is no simple connection in general between the $P_k$s that may be adopted to validate single-channel gains $\gamma_k$ and existence of a solution $P$ that might be used to validate the family of gains $\sqrt{R} \gamma_k$, according to: 
	\begin{equation}
		\label{bigLMIbis}
		\left [ \begin{array}{ccccc}   A_i^T P + P A_i + I & P B_1^i & P B_2^i & \ldots & P B_R^i \\ {B_1^i}^T P & - R \gamma_1^2 I_{m_1} & 0 & \ldots & 0 \\ {B_2^i}^T P & 0 & - R \gamma_2^2 I_{m_2} & \ddots & \vdots \\
			\vdots & \vdots & \ddots & \ddots & 0 \\ {B_R^i}^T P & 0 & \ldots& 0 & - R \gamma_R^2 I_{m_R} \end{array} \right ] \leq 0, \qquad 1 \leq i \leq Q.
	\end{equation}
\end{rem}
\section{The modular small gain condition}
\label{sec:SGlargescale}
To formulate a modular small gain condition we define $3$ types of gains, which we then arrange into suitable matrices.
In particular, for $i<j$, and $k \notin \{i,j\}$
we define $\gamma^{ij}_{kj}$ as the $L_2$ gain of the system:
\begin{equation}
	\dot{\delta} = (A_{ii} \oplus A_{jj}) \delta + H^{ij}_{kj} u.    
\end{equation}
Similarly, for $i<j$ and $k \notin \{i,j\}$ we define
$\gamma^{ij}_{ki}$ as the $L_2$ gain of the system:
\begin{equation}
	\dot{\delta} = (A_{ii} \oplus A_{jj}) \delta + H^{ij}_{ki} u.    
\end{equation}
Moreover, for $1 \leq i \leq N$ and $k \neq i$ we define as $\delta^i_{k}$ the $L_2$ gain of the following system:
\begin{equation}
	\dot{\delta} = A_{ii}^{[2]}  \delta + B_{ik} u.    
\end{equation}
Finally, for $i<j$ we define $\eta_{ij}^1$ and $\eta_{ij}^2$ as the $L_2$ gains of the following systems:
\begin{equation}
	\dot{\delta} = (A_{ii} \oplus A_{jj}) \delta + G_{ij}^1 u,    
\end{equation}
and, respectively, 
\begin{equation}
	\dot{\delta} = (A_{ii} \oplus A_{jj})  \delta + G_{ij}^2 u.    
\end{equation}
We now arrange the gains inside appropriate matrices.
We first define a square matrix $\Gamma$, whose entries are indexed by pairs of integers $i<j$, listed in lexicographical order.
In particular, $(1,2),(1,3),\ldots,(1,N),(2,3), \ldots,(2,N),\ldots, (N-1,N)$.
To this end, for any $i<j$ and $l<m$ we define
\[ [\Gamma]_{ij,lm} = \left \{  \begin{array}{cl} 0 & \textrm{if } (i,j)=(l,m) \textrm{ or } \{i,j\} \cap \{l,m \} =\emptyset \\
	R_{ij} (\gamma^{ij}_{ki})^2 & \textrm{if } i \in \{l,m \} \\
	R_{ij} (\gamma^{ij}_{kj})^2 & \textrm{if } j \in \{l,m\}
\end{array}
\right .
\]
where
\[ R_{ij} = \textrm{card} (\{ (l,k): H^{ij}_{lk} \neq 0 \}) + \textrm{card} (\{ q \in \{1,2\}: G_{ij}^q \neq 0 \}). \]
The matrix $\Delta$ is a rectangular matrix of dimension $N \times { N \choose 2}$ defined as follows:
for all $1 \leq i \leq N$ and all $l<m$ we let:
\[ [\Delta]_{i,lm} = \left \{  \begin{array}{cl} 0 & \textrm{if } i \notin \{l,m \}  \\
	R_{i} (\delta^i_m)^2 & \textrm{if } l=i \\
	R_i (\delta^i_l)^2 & \textrm{if } m=i 
\end{array}
\right .\]
where 
\[R_i = \textrm{card} (\{ k: B_{ik} \neq 0 \}). \]
The matrix $\Upsilon$ is a rectangular matrix of dimension ${ N \choose 2} \times N$ defined as follows:
for all $i<j$ and all $1 \leq k \leq N$,
\[ [\Upsilon]_{ij,k} = \left \{  \begin{array}{cl} 0 & \textrm{if } k \notin \{i,j \}  \\
	R_{ij} (\eta^1_{ij})^2 & \textrm{if } i=k \\
	R_{ij} (\eta^2_{ij})^2 & \textrm{if } j=k 
\end{array}
\right .\]
We are now ready to state our main result.
\begin{thm}
	\label{thm:SG-LSnlsys}
	Consider a nonlinear dynamical system:
	\begin{equation}
		\dot{x}(t) = f(x(t)), 
	\end{equation}
	where $x \in \mathbb{R}^n$ and $f(x): \mathbb{R}^n \rightarrow \mathbb{R}^n$ is a $\mathcal{C}^1$ vector field. Assume that the state vector is partitioned as:
	\[ x = [x_1^T, x_2^T, \ldots, x_N^T]^T \]
	and accordingly the Jacobian matrix
	\[ J(x) = \frac{ \partial f}{\partial x} (x) \]
	admits a block partition as in (\ref{matrixpartition}).
	Define the matrix gains $\Upsilon$, $\Gamma$ and $\Delta$ as introduced above. Then, the second additive compound equation
	\[ \dot{ \delta}^{(2)} (t) = J(x(t))^{[2]} \delta^{(2)} (t) \]
	is exponentially contracting, provided the following small gain condition holds:
	\begin{equation}
		\label{matrixsg}
		\rho( \Upsilon \Delta + \Gamma ) < 1,
	\end{equation}
	where $\rho(\cdot)$ denotes the spectral radius of its argument.
\end{thm}
Before proving Theorem \ref{thm:SG-LSnlsys} we introduce the following Lemma which is is useful to interpret the small gain condition (\ref{matrixsg}).
\begin{lem}
	For a non-negative matrix
	\[  G= \left [  \begin{array}{cc} 0 & \Delta \\ \Upsilon & \Gamma     \end{array} \right ]
	\]  
	the following conditions are equivalent:
	\begin{enumerate}
		\item $\rho( \Upsilon \Delta + \Gamma ) < 1$;
		\item $\rho(G)<1$;
		\item $\rho( \Delta ( I - \Gamma)^{-1} \Upsilon ) < 1$ and $\rho( \Gamma) < 1$.
	\end{enumerate}    
\end{lem}
\emph{Proof.}
We show first $2 \Rightarrow 1.$
For a non-negative matrix, the condition $\rho(G)<1$ is equivalent to existence of a positive vector $v$ such that
$Gv < v$, where $<$ denotes componentwise strict inequalities.
Let condition 2. be satisfied. Then, the vector $v$ can be partitioned according to $[v_1^T,v_2^T]^T$ and condition 2. is equivalent to satisfaction of we inequalities:
\begin{equation}
	\label{schurG}
	\Delta v_2 < v_1, \qquad \Upsilon v_1 + \Gamma v_2 < v_2. 
\end{equation}
Consider the following series of inequalities:
\[  ( \Upsilon \Delta + \Gamma ) v_2 = \Upsilon \Delta v_2 + \Gamma v_2 < \Upsilon v_1 + \Gamma v_2 < v_2 \]
This shows that $(\Upsilon \Delta + \Gamma)$ is Schur. \\
Next, we consider the implication $1 \Rightarrow 2$.
Assume that for some $v_2>0$ we have $(\Upsilon \Delta + \Gamma) v_2 < v_2$. Let $v_1:= \Delta v_2 + \varepsilon \textbf{1}$,
for some sufficiently small $\varepsilon>0$ to be assigned later. Notice that $v_1>0$. Moreover, 
we see that: $(\Upsilon v_1 + \Gamma v_2) = (\Upsilon \Delta + \Gamma) v_2 + \varepsilon \textbf{1} < v_2$, where the latest strict inequality holds provided $0 < \varepsilon < \min_{i} [v_{2i} - [(\Upsilon \Delta + \Gamma) v_2]_i]$.  
At the same time: $\Delta v_2 =  v_1 - \varepsilon \textbf{1} < v_1$.
Hence, we have established inequalities (\ref{schurG}) which imply $G$ is Schur, as previously remarked. \\
Next, we look at the implication $2 \Rightarrow 3$. Consider inequalities (\ref{schurG}). Premultiplication by $\Gamma^k$ of the second inequality yields:
\[  \Gamma^{k} \Upsilon v_1 + \Gamma^{k+1} v_2 < \Gamma^k v_2. \]
Adding these inequalities for $k=0 \ldots K-1$ yields, after getting rid of equal terms on both sides:
\[ \left ( \sum_{k=0}^{K-1} \Gamma^k \right ) \Upsilon v_1 + \Gamma^{K} v_2 < v_2. \]
Since $\Gamma$ is Schur (thanks to the inequality $\Gamma v_2 < v_2$), we may let $K \rightarrow + \infty$ and get:
\[   (I - \Gamma)^{-1} \Upsilon v_1 \leq v_2. \]
Finally, premultiplication times $\Delta$ yields:
\[ \Delta (I- \Gamma)^{-1} \Upsilon v_1 \leq \Delta v_2 < v_1, \]
which proves $\rho ( \Delta (I- \Gamma)^{-1} \Upsilon )<1$. \\
Conversely, let $\rho ( \Delta (I- \Gamma)^{-1} \Upsilon )<1$ and $v_1 >0$ be such that 
$\Delta (I- \Gamma)^{-1} \Upsilon v_1  < v_1$. The condition $\rho( \Gamma )<1$ implies existence
of $\tilde{v}_2>0$ such that $\Gamma \tilde{v}_2 < \tilde{v}_2$.
We let $v_2 = (I- \Gamma)^{-1} \Upsilon v_1 + \varepsilon \tilde{v}_2$, for $\varepsilon>0$ sufficiently small so as to preserve the inequality $\Delta v_2 = \Delta (I- \Gamma)^{-1} \Upsilon v_1 + \varepsilon \Delta \tilde{v}_2 < v_1$.
Remark that:
\[ \Upsilon v_1 + \Gamma v_2 = \Upsilon v_1 + \Gamma (I- \Gamma)^{-1} \Upsilon v_1 + \varepsilon \Gamma  \tilde{v}_2 \]
\[ \qquad =  (I-\Gamma) (I- \Gamma)^{-1} \Upsilon v_1 + \Gamma (I- \Gamma)^{-1} \Upsilon v_1 + \varepsilon \Gamma \tilde{v}_2 \]
\[ \qquad \qquad =   [ (I- \Gamma) + \Gamma) ] (I- \Gamma)^{-1} \Upsilon v_1 + \varepsilon \Gamma \tilde{v}_2  \]
\[ \qquad \qquad = (I-\Gamma)^{-1} \Upsilon v_1 + \varepsilon \Gamma \tilde{v}_2 
< v_2. \]
Hence, the matrix $G$ is Schur. \\

To prove the small gain condition we need to show the connection between matrix $G$, previously defined, and a bounding inequality for solutions of (\ref{diagonalblocks}) and (\ref{offdiagonalblocks}).
The next Lemma provides such a link: \\

\begin{lem}
	Let $X(0)$ be an arbitrary skew symmetric matrix and consider the associated solution of equations (\ref{diagonalblocks}) and (\ref{offdiagonalblocks}) for arbitrary (possibly time-varying) matrices $A(t)$ within the considered domain.
	Denote by $\zeta$ the following vector:
	\begin{equation}
		\zeta = \left[ \| \vec{X}_{11} \|_2^2, 
		\ldots, \| \vec{X}_{NN} \|_2^{2}, \| \textrm{vec} (X_{12}) \|_2^2,
		\|  \textrm{vec} ( X_{13} ) \|_2^2, \ldots, \| \textrm{vec}(X_{(N-1) N} ) \|_2^2 \right]^T. 
	\end{equation}
	The following inequality holds:
	\begin{equation}
		\label{singleinequality}
		\zeta \leq G \zeta + L   |X(0)|^2  \, \textbf{1},  
	\end{equation}
	where $L$ is a sufficiently large constant which can be chosen independently of $X(0)$ and $\textbf{1}$ is the vector of all
	ones of dimension ${{N +1} \choose 2}$. 
\end{lem}
\emph{Proof.}
The result follows by separately considering the case of diagonal ($\vec{X}_{ii}$) and off-diagonal variables $\textrm{vec}(X_{ij})$. 
By equation (\ref{diagonalblocks}), the evolution of each diagonal block $\vec{X}_{ii}$ can be interpreted as a linear (possibly time-varying system) forced by off-diagonal variables. Matrix $\Delta$ is defined on the basis of the $L_2$ gain of individual signals entering each equation and rescaled by the number of simultaneous input, according to the results of Lemma \ref{gainrescale}.
As a consequence, there exists $L$ sufficiently large such that, for all $X(0)$ the following inequality is fulfilled:
\[   \left[ \ldots, \| X_{ii} \|_2^2, \ldots \right]^T \leq \Delta \left[ \| \textrm{vec}(X_{12}) \|_2^2, \ldots,  
\| \textrm{vec} ( X_{(N-1)N} ) \|_2^2 \right]^T + L |X(0)|^2 \textbf{1}.
\]
On the other hand, off-diagonal variables $\textrm{vec} (X_{ij})$ are coupled both with diagonal and off-diagonal ones,
as detailed in (\ref{offdiagonalblocks}). The matrices $\Upsilon$ and $\Gamma$ are also obtained from individual $L_2$ gains, rescaled by the total number of input of each variable, according to the results of Lemma \ref{gainrescale}.
As a consequence, for $L$ sufficiently large, the following inequality holds:
\[   \left[ \| \textrm{vec}(X_{12}) \|_2^2, \ldots,  
\| X_{(N-1)N} \|_2^2 \right]^T \leq \Upsilon \, \left[ \ldots, \| X_{ii} \|_2^2, \ldots \right]^T \]
\[ \qquad \qquad + \Gamma \left[ \| \textrm{vec}(X_{12}) \|_2^2, \ldots,  
\| \textrm{vec} ( X_{(N-1)N} ) \|_2^2 \right]^T + L |X(0)|^2 \textbf{1}.
\]
The combination of the latter two inequalities, in vector form, yields (\ref{singleinequality}) and concludes the proof of the Lemma.
\section{Proof of the main result}
\label{sec:proofMainRes}
We show next how to take advantage of inequality (\ref{singleinequality}) in order to prove convergence and stability of the considered linear system.
To this end, we remark that (\ref{singleinequality}) cannot be used directly to establish boundedness and convergence of the state-variables since its right-hand side involves again $\zeta$, which a priori might be unbounded.
This circularity can be avoided by defining a vector:
\begin{align*}
	 \zeta(t) :=  \bigl[ \| \vec{X}_{11} |_{[0,t]} \|_2^2, 
\ldots, \| \vec{X}_{NN}|_{[0,t]} \|_2^{2},& \| \textrm{vec} (X_{12}) |_{[0,t]} \|_2^2,
\\&\ldots, \| \textrm{vec}(X_{(N-1) N} |_{[0,t]} ) \|_2^2 \bigr]^T. 
\end{align*}
which only considers the $L_2$ norm of signals restricted over the interval $[0,t]$, for each $t \geq 0$.
Due to the linear nature of the system $\zeta(t)$ is well-defined (finite) for each $t \geq 0$.
Moreover, by causality of solutions, it is also true that the following inequality holds:
\begin{equation}
	\label{singleinequalityatt}
	\zeta(t) \leq G \zeta(t) + L   |X(0)|^2  \, \textbf{1},  \qquad \forall \, t \geq 0.
\end{equation}
Notice that $L$ is independent of $t$.
By induction on $k$ one can prove
\[  \zeta(t) \leq G^k \zeta(t) + L |X(0)|^2 ( I + G + \ldots + G^{k-1} ) \textbf{1}. \]
In particular, letting $k \rightarrow + \infty$ in both sides of the above equation yields:
\[ \zeta (t) \leq \lim_{k \rightarrow + \infty} G^k \zeta(t) + L |X(0)|^2 \left ( \sum_{k=0}^{+\infty} G^k \right ) \textbf{1} \]
\[ \qquad \qquad =  L |X(0)|^2  ( I-G )^{-1} \textbf{1} \]
Notice that the right-hand side of the previous inequality is independent of $t$. hence:
\[   \zeta = \lim_{t \rightarrow + \infty} \zeta(t) \leq L |X(0)|^2  ( I-G )^{-1} \textbf{1}.\]
This shows that solutions have bounded $L_2$ norm, and therefore they are bounded (in the $L_{\infty}$ norm) and converge to the origin (given their uniform Lipschitzness over compact sets).

\section{Case study: interconnected Thomas' systems}
\label{sec:example}

The Thomas' system has been introduced by Ren\'e Thomas at the end of the last century as a model capable to reproduce a large class of autocatalytic models that occur in chemical reactions, ecology and evolution (see \cite{sprott2007labyrinth} and references therein). It is described by the following cyclically symmetric first order differential equations:
\begin{equation}
	\label{eq:thomas3}
	\begin{array}{ccc}
		\dot{x}_1 &=& - b x_1 + \sin(x_2) \\
		\dot{x}_2 &=& - b x_2 + \sin(x_3) \\
		\dot{x}_3 &=& - b x_3 + \sin(x_1) 
	\end{array}~,
\end{equation}
where $b$ is a positive parameter. It is well known that system (\ref{eq:thomas3}) admits a forward invariant set of the following form
\begin{equation}
	\label{eq:invset thomas}
	\mathcal{D} := \{x \in \mathbb{R}^3 : b\|x\|_\infty \leq 1\},
\end{equation} 
within which it displays a rich dynamical behavior. Indeed, for $b>1$ it has a unique globally asymptotically stable equilibrium point at the origin, which for $b=1$ undergoes to a supercritical pitchfork bifurcation with the consequent birth of a two stable equilibrium points. 
As $b$ decrease, the system exhibits convergenge towards the equilibrium points until a supercritical Hopf bifurcation occurs at $b=0.32$. Then, for lower values of $b$ complex dynamical behaviors are displayed \cite{sprott2007labyrinth}. In \cite{martini2023bounding} it has been shown that the method introduced in \cite{IEEEmartini} ensures that if $b > 0.442$  the presence of attractors with positive Lyapunov exponents is excluded inside the invariant set (\ref{eq:InvariantSet}), while in \cite{angeli2023small}, where the system is considered as the interconnection of two different subsystems, we get the more conservative bound $b > 0.575$.\\
In the present paper, we consider the system $\Sigma$ composed by three interconnected Thomas' systems in the following setting:
\begin{align}
	\Sigma:\begin{cases}\Sigma_1:\begin{cases}\begin{array}{ccl}
				\dot{x}_1 &=& - b\,x_1 + \sin(x_2) -d\,x_4 \\
				\dot{x}_2 &=& - b\,x_2 + \sin(x_3) \\
				\dot{x}_3 &=& - b\,x_3 + \sin(x_1) 
		\end{array}\end{cases}\\\label{eq:ThomasInter}
		\Sigma_2:\begin{cases}\begin{array}{ccl}
				\dot{x}_4 &=& - a\,x_4 + \sin(x_5) -d\,x_7\\
				\dot{x}_5 &=& - a\,x_5 + \sin(x_6) \\
				\dot{x}_6 &=& - a\,x_6 + \sin(x_4) 
		\end{array}\end{cases}\\
		\Sigma_3:\begin{cases}\begin{array}{ccl}
				\dot{x}_7 &=& - a\,x_7 + \sin(x_8) -d\,x_1\\
				\dot{x}_8 &=& - a\,x_8 + \sin(x_9) \\
				\dot{x}_9 &=& - a\,x_9 + \sin(x_7) 
		\end{array}\end{cases}
	\end{cases},
\end{align}
where $a=2$, $d$ represents the coupling strength, $d\in [-1,1]$, and $b$ is a positive parameter. Clearly, when $d = 0$, system $\Sigma$ is composed of three decoupled Thomas’ systems which can be analyzed separately according to \cite{martini2023bounding} and \cite{angeli2023small}. \\
When $d \neq 0$, the approach in \cite{martini2023bounding} directly applied to the $2$-additive compound of system's Jacobian of dimension $36 \times 36$ might be computationally awkward. Instead, through the method introduced in Section \ref{sec:SGlargescale}, it can be verified that the maximum dimension of the matrix to be handled is $9\times 9$, corresponding to the matrices $H_{kj}^{ij}$, $H_{ki}^{ij}$ and those given by the Kronecker sum of the subsystems' Jacobian $(J_{ii} \oplus J_{jj})$. \\
The Jacobian of the system reads
\begin{align}
	\label{eq:Jacobian}
	J = \left(\begin{array}{ccccccccc} -b & c_2 & 0 & -d & 0 & 0 & 0 & 0 & 0\\ 0 & -b & c_3 & 0 & 0 & 0 & 0 & 0 & 0\\ c_1 & 0 & -b & 0 & 0 & 0 & 0 & 0 & 0\\ 0 & 0 & 0 & -a & c_5 & 0 & -d & 0 & 0\\ 0 & 0 & 0 & 0 & -a & c_6 & 0 & 0 & 0\\ 0 & 0 & 0 & c_4 & 0 & -a & 0 & 0 & 0\\ -d & 0 & 0 & 0 & 0 & 0 & -a & c_8 & 0\\ 0 & 0 & 0 & 0 & 0 & 0 & 0 & -a & c_9\\ 0 & 0 & 0 & 0 & 0 & 0 & c_7 & 0 & -a \end{array}\right),
\end{align}
where $c_i = \cos{x_i}$, $i=1,\ldots,9$. Therefore, for example, we have that the $2$-additive compound of subsystems $\Sigma_1$, named $J_{11}^{[2]}$, assumes the form
\begin{equation}
	J_{11}^{[2]}=\left(\begin{array}{ccc} -2\,b & c_3 & 0\\ 0 & -2\,b & c_2\\ -c_1 & 0 & -2\,b \end{array}\right),
\end{equation}
while, the Kronecker sum of the subsystems $\Sigma_1$ and $\Sigma_2$ is equal to
\begin{align}
	(J_{11} \oplus J_{22}) = \left(\begin{array}{ccccccccc} c & c_5 & 0 & c_2 & 0 & 0 & 0 & 0 & 0\\ 0 & c & c_6 & 0 & c_2 & 0 & 0 & 0 & 0\\ c_4 & 0 & c & 0 & 0 & c_2 & 0 & 0 & 0\\ 0 & 0 & 0 & c & c_5 & 0 & c_3 & 0 & 0\\ 0 & 0 & 0 & 0 & c & c_6 & 0 & c_3 & 0
		\\ 0 & 0 & 0 & c_4 & 0 & c & 0 & 0 & c_3\\ c_1 & 0 & 0 & 0 & 0 & 0 & c & c_5 & 0\\ 0 & c_1 & 0 & 0 & 0 & 0 & 0 & c & c_6\\ 0 & 0 & c_1 & 0 & 0 & 0 & c_4 & 0 & c \end{array}\right),
\end{align}
where $c = -(b+a)$. 
In the appendix it is shown that the hyper rectangle in (\ref{eq:InvariantSet}) is a forward invariant sets of (\ref{eq:ThomasInter}), where $X_1$, $X_4$, $X_7$ are as in (\ref{eq:X1X4X7}). As a consequence, the $2$-additive compound of the Jacobian of each subsystem belongs to the corresponding convex hull ${\rm conv}(\{J_{ii,1}^{[2]},\dots,J_{ii,8}^{[2]}\})$ of vertices given by all the combinations of the upper and lower bounds:
\begin{equation}
	\label{eq:ConvexHulls}
	\begin{array}{cc}
			c_1 \,\in\, \left[\cos{X_1},1\right],  &
			c_2\,,\,c_3\in\left[\cos{\dfrac{1}{b}},1\right]\vspace{0.2cm}\\
			c_4 \,\in\, \left[\cos{X_4},1\right],  &
			c_5\,,\,c_6\in\left[\cos{\dfrac{1}{a}},1\right]\vspace{0.2cm}\\
			c_7 \,\in\, \left[\cos{X_7},1\right] , &
			c_8\,,\,c_9\in\left[\cos{\dfrac{1}{a}},1\right]
		\end{array} .
\end{equation}
Moreover, the Kronecker sums $(J_{ii} \oplus J_{jj})$ of the subsystems belong to the convex hull ${\rm conv}(\{(J_{ii} \oplus J_{jj})_1,\dots,(J_{ii} \oplus J_{jj})_{64}\})$ of vertices given by all the combinations of the upper and lower bounds listed in (\ref{eq:ConvexHulls}). Similar considerations can be also made for matrices $B_{ik}$, $G^1_{ij}$, $G^2_{ij}$, $H^{ij}_{kj}$ and $H^{ij}_{ki}$, obtaining the corresponding convex hulls. It is worth pointing out that the centralized approach to the $2$-additive compound of the system Jacobian of dimension $36\times 36$ would require evaluation of $J$ on $2^9=512$ vertices, instead. 

\begin{figure}[h!]
	\begin{center}
		\begin{tabular}{cc}
			\includegraphics[width=0.45\textwidth]{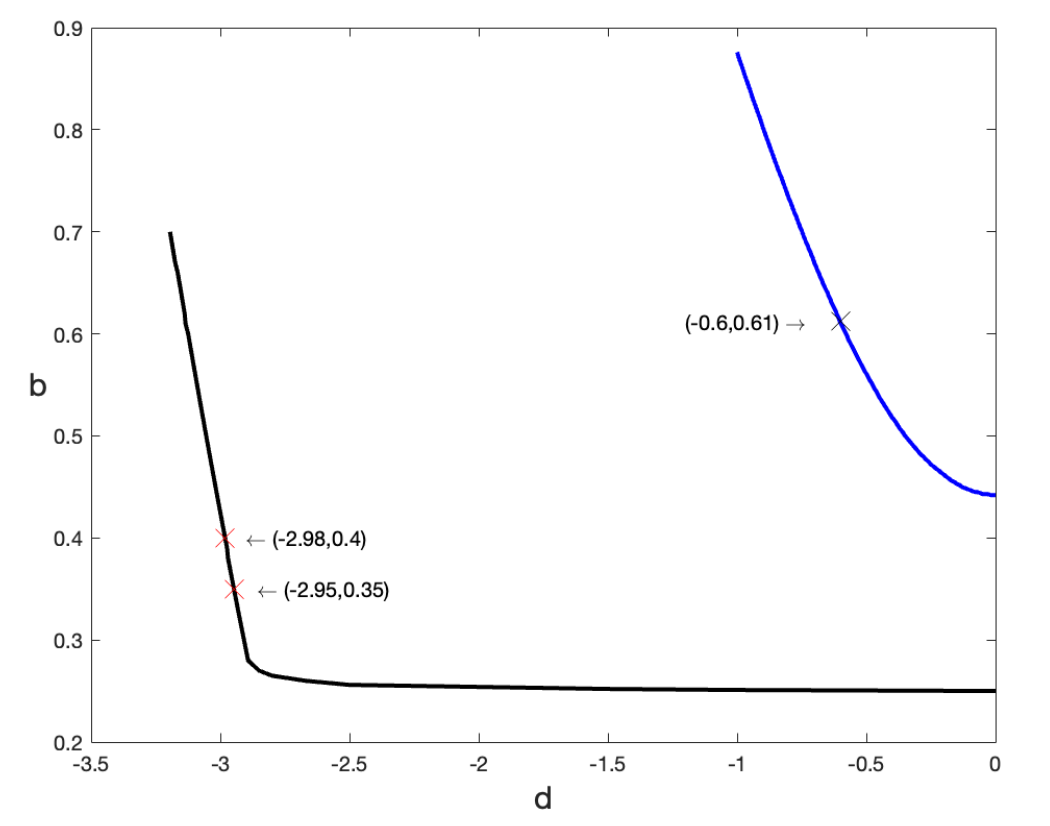} &
			\includegraphics[width=0.45\textwidth]{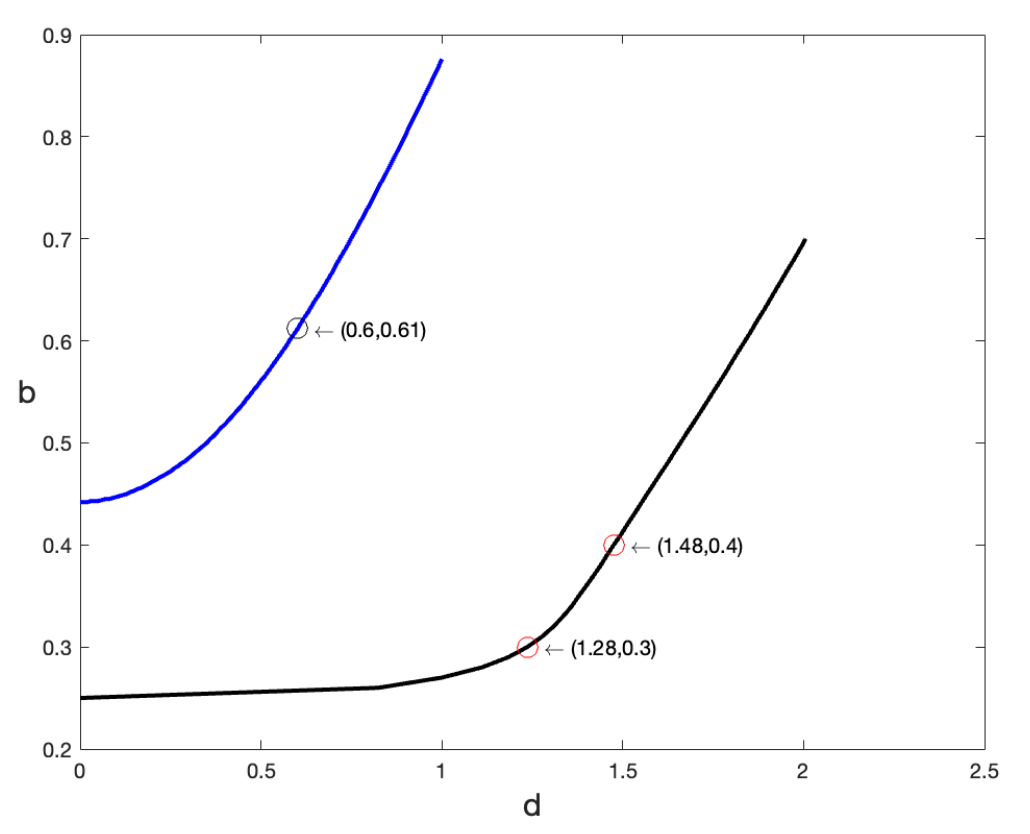}\\
			a) & b) 
		\end{tabular}
	\end{center}
	\caption{a) Parameter $b$ as a function of the coupling strength $d$, $d \in [-1,0]$; b) Parameter $b$ as a function of the coupling strength $d$, $d \in [0,1]$. The blue curve is obtained with the small-gain like approach, while the black curve is obtained with the root locus approach.}
	\label{fig:B_D}
\end{figure}

Our aim is to find, for each value of the coupling strength $d$ within the interval $[-1,1]$, the value of the parameter $b$ of subsystem $\Sigma_1$ for which condition (\ref{matrixsg}) is satisfied and the $2$-additive compound of system's Jacobian (\ref{eq:Jacobian}) is exponentially contracting, implying convergence towards equilibria for solutions of system $\Sigma$.\\ Therefore, to verify condition (\ref{matrixsg}) we look for the gains $\delta_k^i$ solving the following minimization problems
\begin{align} 
	\label{SG_lmi_deltagains}
	&\qquad\qquad\qquad\qquad ~ ~\min_{b \geq 0, P_{ik} = P_{ik}^T }  b \notag\\
	& \qquad\qquad\qquad\qquad\qquad\textrm{subject to} \notag\\
	& \left[\begin{array}{cc}
		{J_{ii,q}^{[2]}}^T P_{ik} +P_{ik} J_{ii,q}^{[2]}+ I & P_{ik} B_{ik} \\
		B_{ik}^T P_{ik} & - {\delta_{k}^i}^2 I_9
	\end{array}\right] \leq 0, \, q=1,\ldots,8 ,\\ 
	& \qquad\qquad\qquad\qquad\qquad ~~P_{ik} \geq  0 \notag\\
	&\qquad\qquad\qquad\qquad\qquad ~~\delta_{k}^i \geq 0 \notag
\end{align}
where the matrix $B_{ik}$ is the corresponding matrix in (\ref{eq:Bik}) to the considered gain $\delta_k^i$. While, for the gains $\eta_{ij}^1$ and $\eta_{ij}^2$ we have
\begin{align} 
	\label{SG_lmi_etagains}
	&\qquad\qquad\qquad\qquad\qquad\quad ~\min_{b \geq 0, P_{ik}^l = {P_{ik}^l}^T }  b \notag\\
	& \qquad\qquad\qquad\qquad\qquad\qquad \quad\textrm{subject to} \notag\\
	& \left[\begin{array}{cc}
		(J_{11} \oplus J_{22})_q^T P_{ik}^l +P_{ik}^l (J_{11} \oplus J_{22})_q+ I & P_{ik}^l G_{ij}^l \\
		{G_{ij}^l}^T P_{ik}^l & - {\eta_{ij}^l}^2 I_3
	\end{array}\right] \leq 0, \, q=1,\ldots,64 ,\\ 
	&\qquad\qquad\qquad\qquad\qquad\qquad\quad ~~ P_{ik}^l \geq  0 \notag\\
	&\qquad\qquad\qquad\qquad\qquad\qquad\quad ~~ \eta_{ij}^l \geq 0 \notag
\end{align}
where $G_{ij}^l$ are the matrices in (\ref{eq:Gik}) corresponding to the gain $\eta_{ij}^l$, with $l=1,2$. Finally, for the gains $\gamma_{kj}^{ij}$ and $\gamma_{ki}^{ij}$, the minimization problem boils down to
\begin{align} 
	\label{SG_lmi_gammagains}
	&\qquad\qquad\qquad\qquad\qquad\qquad \min_{b \geq 0, P_l^{ij} = {P_l^{ij}}^T }  b \notag\\
	&\qquad\qquad\qquad\qquad\qquad\qquad \quad~~\textrm{subject to} \notag\\
	& \left[\begin{array}{cc}
		(J_{11} \oplus J_{22})_q^T P_l^{ij} +P_l^{ij} (J_{11} \oplus J_{22})_q+ I & P_l^{ij} H_l^{ij} \\
		{H_l^{ij}}^T P_l^{ij} & - {\gamma_l^{ij}}^2 I_9
	\end{array}\right] \leq 0, \, q=1,\ldots,64 ,\\ 
	&\qquad\qquad\qquad\qquad\qquad\qquad\quad\quad ~P_l^{ij} \geq  0 \notag\\
	&\qquad\qquad\qquad\qquad\qquad\qquad\quad\quad ~\gamma_l^{ij} \geq 0 \notag
\end{align}
where $H_l^{ij}$ are the matrices in (\ref{eq:Gik}) corresponding to the gain $\gamma_{l}^{ij}$, with $l=(kj),(ki)$. \\
Once all the above gains have been computed for given values of $b$ and $d$, condition (\ref{matrixsg}) can be verified. The blue curve in Fig. \ref{fig:B_D} denotes for each fixed $d$ the value of $b$ under which condition (\ref{matrixsg}) is not verified anymore. In order to understand the conservatism of the approach, we exploit tha fact that the $2$-additive compound of the Jacobian, computed on the equilibrium points which are inside the invariant set of the system, has to be marginally stable \cite{martini2023bounding}. For the sake of simplicity, we apply this necessary condition only to the equilibrium points in $x=0$. Hence, considering the system's Jacobian for $x=0$ and, recalling the spectral properties of additive compound matrices, we can find the curve on the $(d,b)$ plane for the marginal stability of $J^{[2]}(0)$ computing the values of $d$ and $b$ for which the sum of the real part of different eigenvalues of the matrix $J(0)$ is equal to zero. \\
The system's Jacobian computed in $x=0$ reads
\begin{equation}
	J = \left(\begin{array}{ccccccccc} -b & 1 & 0 & -d & 0 & 0 & 0 & 0 & 0\\ 0 & -b & 1 & 0 & 0 & 0 & 0 & 0 & 0\\ 1 & 0 & -b & 0 & 0 & 0 & 0 & 0 & 0\\ 0 & 0 & 0 & -a & 1 & 0 & -d & 0 & 0\\ 0 & 0 & 0 & 0 & -a & 1 & 0 & 0 & 0\\ 0 & 0 & 0 & 1 & 0 & -a & 0 & 0 & 0\\ -d & 0 & 0 & 0 & 0 & 0 & -a & 1 & 0\\ 0 & 0 & 0 & 0 & 0 & 0 & 0 & -a & 1\\ 0 & 0 & 0 & 0 & 0 & 0 & 1 & 0 & -a \end{array}\right) = \left[\begin{array}{cc}
		J_{11}(0)  & L \\
		L^\top  & M
	\end{array}\right],
\end{equation}
where
\begin{align*}
	& J_{11}(0) = \left[\begin{array}{ccc}
		-b & 1 & 0   \\
		0 & -b & 1 \\
		1 & 0 & -b
	\end{array}\right],\\
	&L = \left[\begin{array}{cccccc}
		-d & 0 & 0 & 0 & 0 & 0 \\
		0 & 0 & 0 & 0 & 0 & 0\\
		0 & 0 & 0 & 0 & 0 & 0
	\end{array}\right],\\
	&M = \left[\begin{array}{cccccc}
		&  &  & -d & 0 & 0 \\
		& J_{22}(0) &  & 0 & 0 & 0\\
		&  &  & 0 & 0 & 0\\
		0 & 0 & 0 &  &  & \\
		0 & 0 & 0 &  &J_{33}(0)  & \\
		0 & 0 & 0 &  &  & 
	\end{array}\right].
\end{align*}
and
\begin{equation*}
	J_{22}(0) = J_{33}(0) = \left[\begin{array}{ccc}
		-a & 1 & 0   \\
		0 & -a & 1 \\
		1 & 0 & -a
	\end{array}\right].
\end{equation*}
Exploiting the Schur's complement, it can be shown that the determinant of $J(0)$ can be written as
\begin{equation}
	\det(J(0)) = \det(J_{11}(0))\,\det(J_{22}(0))\,\det(J_{33}(0))-d^3\,b^2\,a^4,
\end{equation}
and, as a consequence, it turns out that the characteristic polynomial of $J(0)$ assumes the form
\begin{equation}
	\label{eq:detQ}
	\det(\lambda I_9 - J(0)) = \prod\limits_{i=1}^3\det(\lambda I_3-J_{ii}(0))-d^3(\lambda - b)^2\,(\lambda - a)^4 .
\end{equation}
Equation (\ref{eq:detQ}) can be interpreted as the root locus of the transfer function 
\begin{equation}
	G(s) = \dfrac{\prod\limits_{i=1}^3\det(s\,I_3-J_{ii}(0))}{(s - b)^2\,(s - a)^4},
\end{equation}
with gain $\gamma = -d^3$. Therefore, it can be used to graphically find a bound for the marginal stability of the matrix $J^{[2]}(0)$, finding, for any given value $b_i$ of the parameter b, the value $\gamma_i$ of minimum magnitude of the gain $\gamma$ for which at least one sum of the real part of different eigenvalues is equal to zero. Therefore, the curve on the $(d,b)$-plane identifying marginal stability of $J^{[2]}(0)$ can be drawn via the points $(\sqrt[3]{\gamma_i},b_i)$.

\begin{figure}[h!]
	\begin{center}
		\begin{tabular}{cc}
			\includegraphics[width=0.5\textwidth]{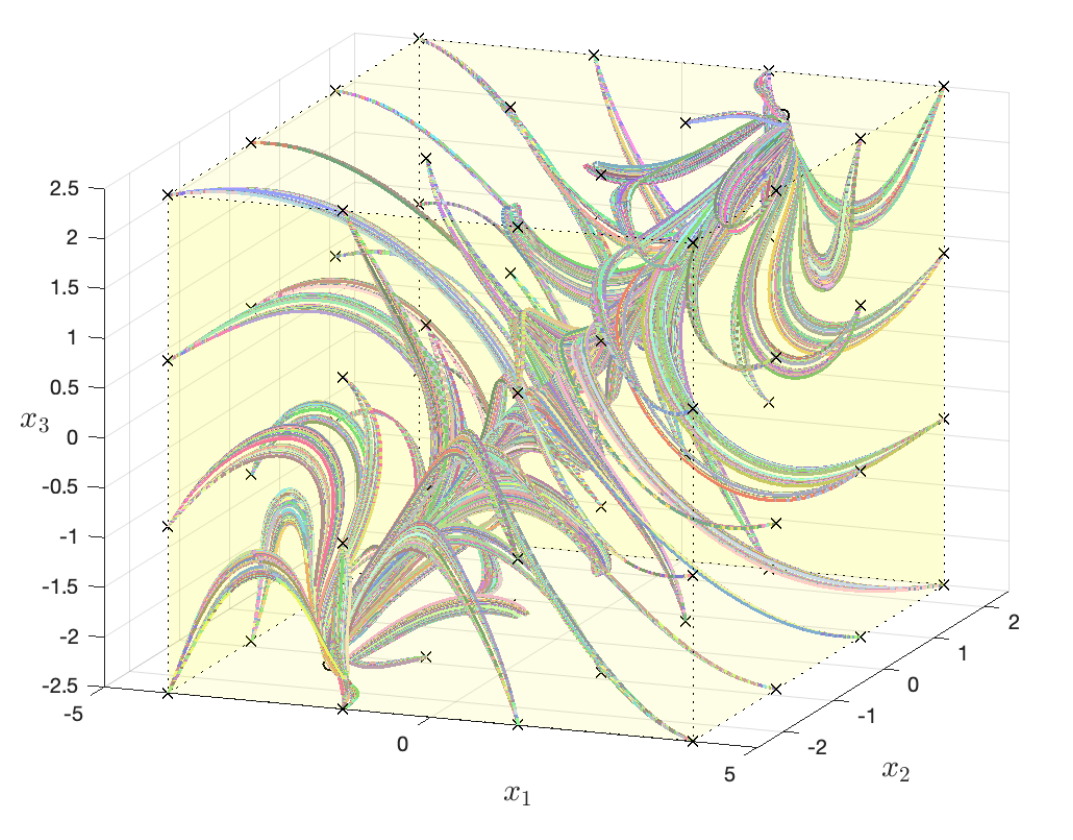} &
			\includegraphics[width=0.5\textwidth]{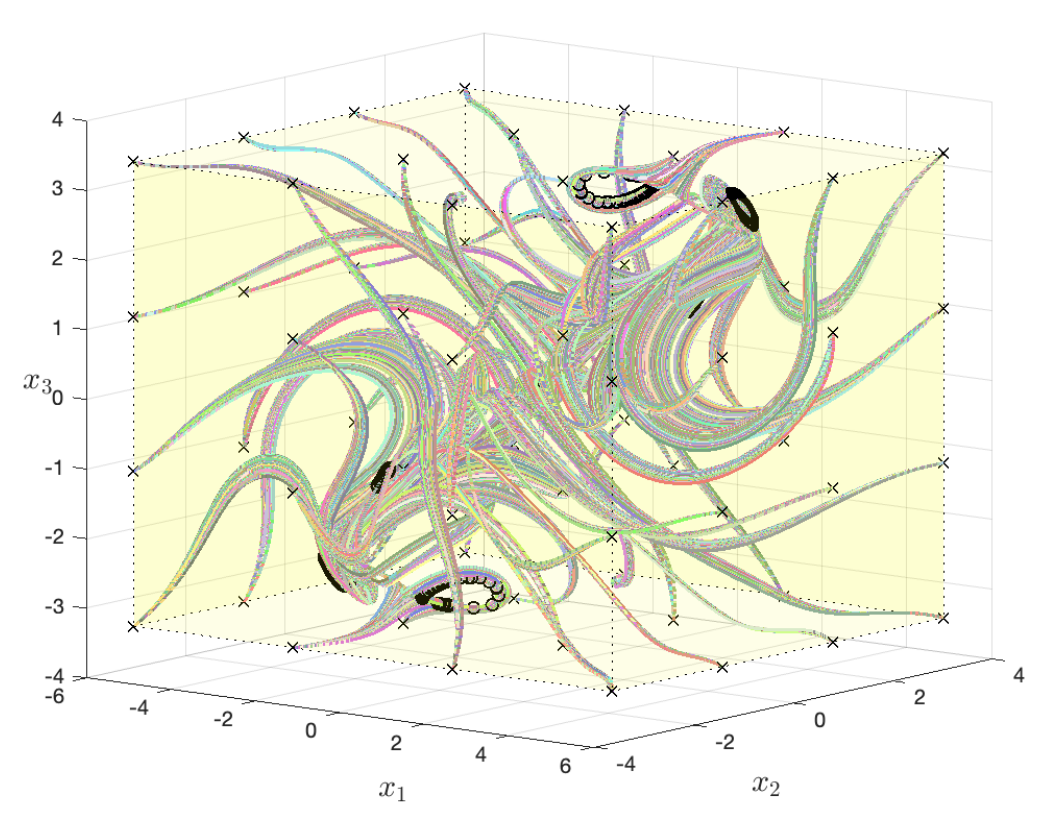}\\
			a) & b) 
		\end{tabular}
	\end{center}
	\caption{Simulation of system $\Sigma$ projected on the $(x_1,x_2,x_3)$ plane for different initial conditions (marked by x). a) $b$ = 0.4, $d$ = 0.6: convergence towards the equilibrium points (marked with o); b) $b$ = 0.3, $d$ = 0.6: presence of multiple stable oscillatory behaviours (marked black).}
	\label{fig:SimDpos}
\end{figure}

\begin{figure}[h!]
	\begin{center}
		\begin{tabular}{cc}
			\includegraphics[width=0.5\textwidth]{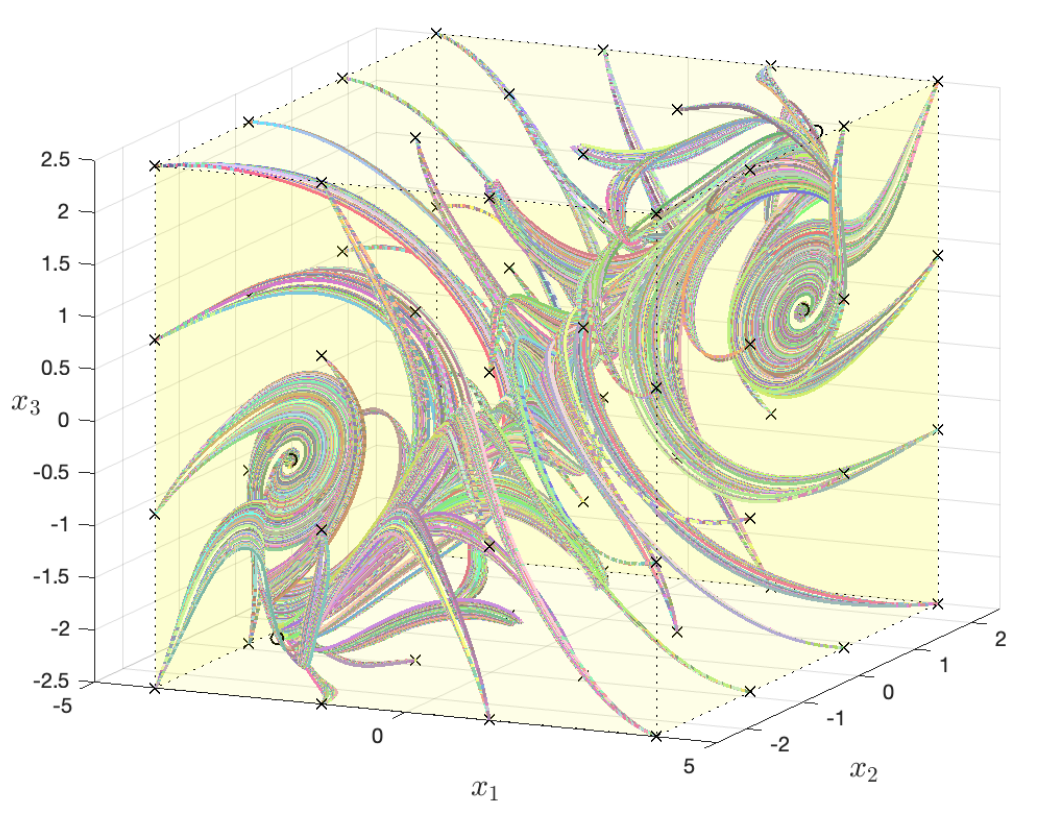} &
			\includegraphics[width=0.5\textwidth]{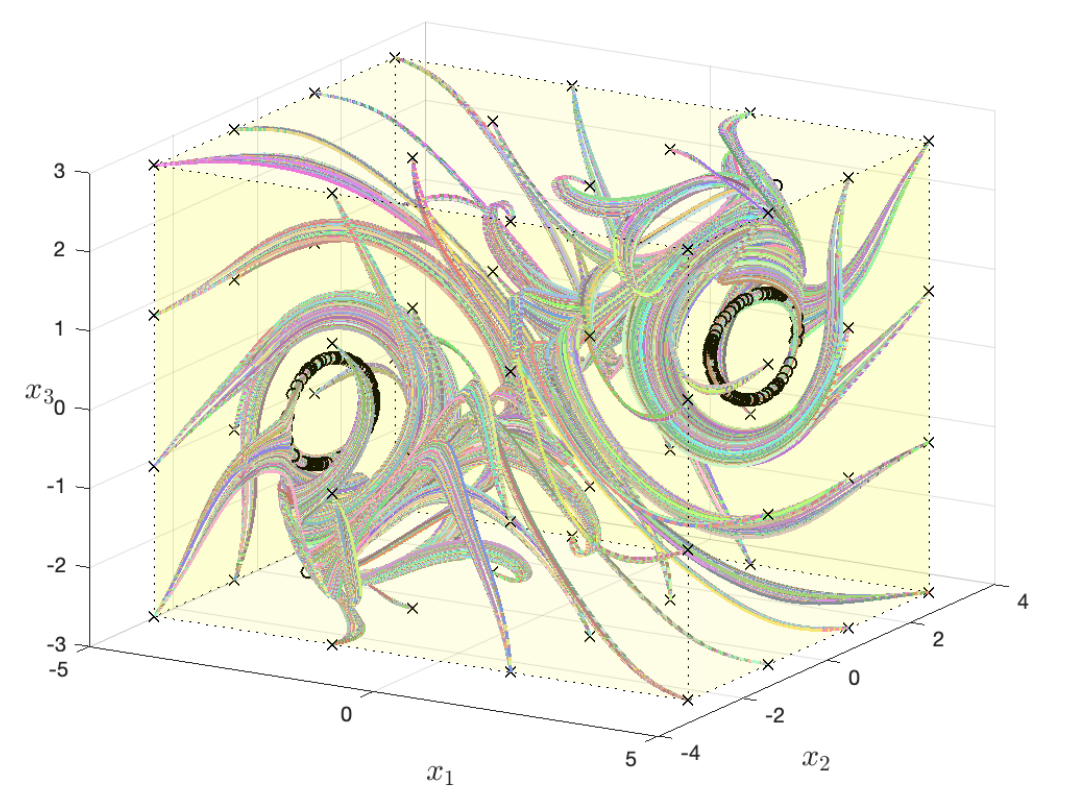}\\
			a) & b) 
		\end{tabular}
	\end{center}
	\caption{Simulation of system $\Sigma$ projected on the $(x_1,x_2,x_3)$ plane for different initial conditions (marked x). a) $b$ = 0.4, $d$ = 0.6: convergence towards the equilibrium points (marked with o); b) $b$ = 0.3, $d$ = 0.6: presence of multiple stable oscillatory behaviours (marked black).}
	\label{fig:SimDneg}
\end{figure}

By proceeding in this way, we obtain the black curve reported in Fig. \ref{fig:B_D}. Note that for $d=0$ we obtain the same result as in \cite{martini2023bounding}. \\
As expected, the blue curve is always greater than the black one, showing that it could be some conservatism in the developed small-gain like approach. 
However, it is worth noting that the approach based on the root locus of $J(0)$ gives us a curve which is in general a lower bound of the curve which separates convergent from oscillatory or more complex dynamics. Therefore, the real conservatism of the small-gain like approach could be less than the one displayed in Fig. \ref{fig:B_D}. 
Indeed, this is highlighted in Figs. \ref{fig:SimDpos} and \ref{fig:SimDneg} where both convergence towards equilibria and oscillations are displayed for values of $b$ lying between the two curves.

\begin{figure}[h!]
	\centering
	\includegraphics[width=0.6\columnwidth]{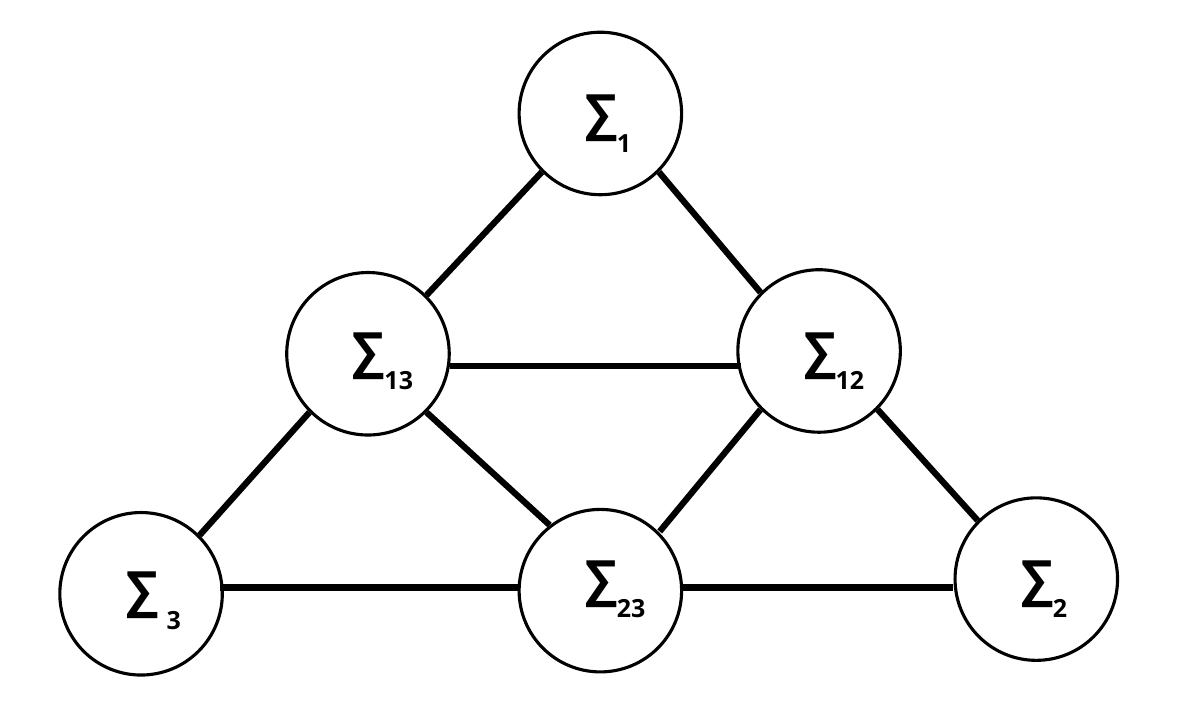}
	\caption{Topology of interconnection of $J^{[2]}$: $\Sigma_i$s denote the ``diagonal'' subsystems
		$X_{ii}$, while $\Sigma_{ij}$ denote the off-diagonal systems $X_{ij}$.}
	\label{fig:Config}
\end{figure}
\section{Conclusion and final remarks}
\label{sec:concl}
This paper proposes a small-gain like condition ensuring $2$-contraction of a large-scale interconnected nonlinear system, which in turn enables convergence of solutions towards the equilibrium points of the overall system.
The exponential contraction of the Jacobian's $2$-additive compound is guaranteed if the spectral radius of the gain matrix $G$ is less than one. It is worth noting that the spectral radius of the matrix $G$ depends on the $L_2$ gains of diagonal $\Sigma_i$ and off-diagonal $\Sigma_{ij}$ subsystems that dictate the second additive dynamics of the interconnection. 
This means that in the case of three subsystems all connected together, we obtain the situation reported in Fig. \ref{fig:Config}. As a consequence, the situation become more complicated as the number of interconnected systems grows, since the number of fictitious systems is equal to $N \choose 2$, with $N$ the number of interconnected systems. However, the dimension of the LMI problems to be solved for computing the gains of the subsystems, is significantly smaller than that employed to directly verify $2$-contraction of the $2$-additive compound matrix of the Jacobian of the overall system.

\appendix
\section{Invariant set for system $\Sigma$}

\begin{prop}
	For the dynamics of system $\Sigma$ in (\ref{eq:ThomasInter}) the hyper rectangle defined as
	\begin{align} 
		\label{eq:InvariantSet}
		\notag    \Biggl\{ x \in \mathbb{R}^9:
		|x_1| \leq X_1,~ |x_2|\leq \dfrac{1}{b}, ~|x_3| &\leq \dfrac{1}{b}, ~|x_4| \leq X_4, ~ |x_5| \leq \dfrac{1}{a},~ |x_6|\leq\dfrac{1}{a}, \\  & ~|x_7| \leq X_7,~ |x_8|\leq\dfrac{1}{a}, ~|x_9| \leq \dfrac{1}{a} \Biggr\}
	\end{align}
	where
	\begin{equation}
		\label{eq:X1X4X7}
		\begin{array}{cc}
			\begin{array}{ccc}
				X_1 &=& \dfrac{\left(1+\dfrac{|d|}{a}+\left(\dfrac{|d|}{a}\right)^2\right)}{b\,\left(1-\dfrac{|d|}{b}\left(\dfrac{|d|}{a}\right)^2\right)}\end{array} \,\,\,,&
			\begin{array}{ccc}
				X_4 &=& \dfrac{\left(1+\dfrac{|d|}{a}+\dfrac{|d|^2}{a\,b}\right)}{a\,\left(1-\dfrac{|d|}{b}\left(\dfrac{|d|}{a}\right)^2\right)} \end{array}\vspace{0.2cm}\,\,\,,\\
			\begin{array}{ccc}
				X_7 &=& \dfrac{\left(1+\dfrac{|d|}{b}+\dfrac{|d|^2}{a\,b}\right)}{b\,\left(1-\dfrac{|d|}{b}\left(\dfrac{|d|}{a}\right)^2\right)} \end{array}\,\,\, .
		\end{array}
	\end{equation}
	is forward invariant.
\end{prop}
{\it Proof.} - A bound on the variables $x_1$, $x_4$ and $x_7$ can be found by solving the following system of equations:
\begin{equation}
	\begin{array}{ccc}
		-b\,X_1 + 1 + |d|\,X_4 & = & 0  \\
		-a\,X_4 + 1 + |d|\,X_7 & = & 0 \\
		-a\,X_7 + 1 + |d|\,X_1 & = & 0
	\end{array},
\end{equation}
whose solution gives
\begin{equation}
	\begin{array}{cc}
		\begin{array}{ccc}
			X_1 &=& \dfrac{\left(1+\dfrac{|d|}{a}+\left(\dfrac{|d|}{a}\right)^2\right)}{b\,\left(1-\dfrac{|d|}{b}\left(\dfrac{|d|}{a}\right)^2\right)}\end{array} \,\,\,,&
		\begin{array}{ccc}
			X_4 &=& \dfrac{\left(1+\dfrac{|d|}{a}+\dfrac{|d|^2}{a\,b}\right)}{a\,\left(1-\dfrac{|d|}{b}\left(\dfrac{|d|}{a}\right)^2\right)} \end{array}\vspace{0.2cm}\,\,\,,\\
		\begin{array}{ccc}
			X_7 &=& \dfrac{\left(1+\dfrac{|d|}{b}+\dfrac{|d|^2}{a\,b}\right)}{b\,\left(1-\dfrac{|d|}{b}\left(\dfrac{|d|}{a}\right)^2\right)} \end{array}\,\,\, .
	\end{array}
\end{equation}
Application of Nagumo's criterion shows that the set 
\[ \{ x \in \mathbb{R}^9:
|x_1| \leq X_1, |x_4| \leq X_4, |x_7| \leq X_7 \} \] is forward invariant.
Moreover, each of the remaining variables can be interpreted as the state of a scalar stable linear system forced by
a bounded disturbance taking values in $[-1,1]$. We may therefore assume bounds of the form $|x_i| \leq X_i$ for
\begin{equation}
	\begin{array}{l}
		X_2\,=\,X_3\,=\,\dfrac{1}{b}\\
		X_5\,=\,X_6\,=\,X_8\,=X_9\,=\,\dfrac{1}{a}
	\end{array}. 
\end{equation}
Combining these bounds with those derived for the variables $x_1$, $x_4$ and $x_7$ yields the desired result.
\qed


\end{document}